\documentclass{PoS}

\usepackage{cite}

\newcommand{\pv}{\mathbf{p}}
\newcommand{\kv}{\mathbf{k}}
\newcommand{\rv}{\mathbf{r}}

\title{Calculation of free baryon spectral densities at finite temperature}

\ShortTitle{Calculation of free baryon spectral densities at finite temperature}

\author{\speaker{Chrisanthi Praki} and Gert Aarts\\
        Department of Physics, College of Science, Swansea University, Swansea, United Kingdom \\
        E-mail:  \email{c.s.praki.520287@swansea.ac.uk}, \email{g.aarts@swan.ac.uk}
        }

\abstract{Following a recent lattice study of nucleon parity doubling at finite temperature from the computation of the two-point nucleon correlators, we study the spectral functions of free nucleons at finite temperature. Spectral densities in the continuum are presented along with a comparison to (free) results on the lattice. Particular attention is given to lattice artefacts at higher energies.
}

\FullConference{The 33rd International Symposium on Lattice Field Theory\\
		14 -18 July 2015\\
		Kobe International Conference Center, Kobe, Japan
		}

\begin{document}

\section{Introduction}

The properties of hadrons may be effected by the presence of a thermal medium. Examples of these effects include e.g.\ a thermal mass shift,
thermal broadening, the melting/dissolution of hadrons and degeneracy due to chiral symmetry restoration  \cite{Rapp:1999ej}.
   For mesons, in-medium effects have been analysed intensely on the lattice, both for light quarks and for heavy quarks (quarkonia), see e.g.\  Ref.\  \cite{Ding:2012ar} for a recent review. 
    However, similar lattice studies involving baryons are rather limited:  screening masses were studied some time ago \cite{DeTar:1987ar, Pushkina:2004wa} and a study of temporal correlators for quenched QCD can be found in Ref.\ \cite{Datta:2012fz}. The first detailed lattice study of nucleons at finite temperature across the deconfinement transition was presented recently in Ref.\ \cite{Aarts:2015mma}, where parity doubling was seen to emerge as the temperature is increased. Moreover, in the confined phase the nucleon ground state appeared to be largely independent of the temperature, whereas temperature effects are substantial in the negative-parity ($N^*$) channel. These results might be relevant for further developing e.g.\ parity-doublet and other models attempting to explain the role of chiral symmetry in the hadronic spectrum     \cite{Detar:1988kn,Benic:2015pia,Weyrich:2015hha,Motohiro:2015taa,Hohler:2015iba}.

The next step in the analysis will be the study of nucleon spectral functions \cite{Allton:lat15}. In order to prepare for that, we study here nucleon spectral densities in the absence of interactions, both in the continuum and on the lattice. A comparison of the two allows us to look at discretisation effects and, in particular, lattice artefacts at higher energies.

\section{Calculation of the free nucleon propagator} 

The free nucleon propagator may be calculated by considering a nucleon composed of three non-interacting quarks produced at a point ($\mathbf{0},0$) and propagating to the point ($\mathbf{x},\tau$) where it annihilates. As the quarks are non-interacting, they are not confined: hence the quantitative results are relevant at very high temperature in the quark-gluon plasma due to asymptotic freedom.
Nevertheless, since we use operators with the same quantum numbers as the nucleon, we will refer to the correlator as the nucleon correlator, etc.

\begin{figure}[h]

\begin{center}

\setlength{\unitlength}{1cm}
\thicklines
\vspace*{20pt}
\begin{picture}(10,4.5)
\put(2,3){\vector(1,0){3}}
\put(5,3){\line(1,0){3}}
\put(2,3){\circle*{0.1}}
\put(8,3){\circle*{0.1}}
\qbezier(2,3)(5, 7)(8, 3)
\qbezier(2,3)(5, -1)(8, 3)
\put(4.9,5){\vector(1,0){0.1}}
\put(4.9,1){\vector(1,0){0.1}}
\put(1.2,3)
{($\mathbf{0}$,0)}
\put(8.1,3)
{($\mathbf{x}$,$\tau$)}
\end{picture}
\end{center}
\vspace*{-30pt}
\caption{A nucleon is produced at point ($\mathbf{0}$,0) and propagates to the point ($\mathbf{x}$,$\tau$) where it  annihilates.}
\label{prop}
\end{figure}
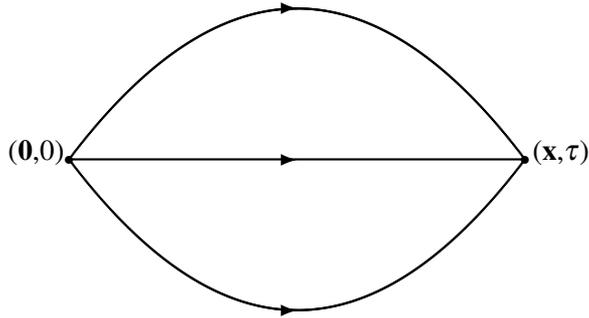

We introduce the proton and antiproton interpolating fields as follows,
\begin{equation}
O^\alpha=\epsilon_{abc}[d_{a}^TC^{-1}\gamma_5u_b]u_c^{\alpha},
\qquad\qquad
\bar{O}^{\alpha'}=\epsilon_{a'b'c'}\bar{u}_{a'}^{\alpha'}[\bar{u}_{b'}\gamma_5C\bar{d}_{c'}^{T}].
\end{equation}
where the Greek letters correspond to the Dirac components, the Latin letters to the colour components and $u$ and $d$ denote the flavour content. The charge conjugation matrix is defined as $C=i\gamma_2\gamma_4$. 

The quark propagator is decomposed as
\begin{equation}
 \label{decomp}
 S_{ab}(x)=\left\langle \psi_{a}(x)\bar{\psi}_{b}(0)\right\rangle = \delta_{ab} \left[\sum_{\nu=1}^4 S_\nu(x)\gamma_\nu+\mathrm{I}_4S_m(x) \right],
\end{equation}
where $\psi = u, d$.
 The two-point correlation function may then be expressed as a combination of gamma matrices and $B$ coefficients, 
\begin{equation}
G^{\alpha \alpha'}(x)=\langle O^\alpha(x)\bar{O}^{\alpha'}(0)\rangle = \sum_\nu B_\nu(x)\gamma_\nu^{\alpha\alpha'}+\mathrm{I}_4^{\alpha\alpha'}B_m(x),
\end{equation}
where the $B$ coefficients are given by  \cite{Cichy:2008gk}
\begin{equation}
B_\mu=6S_\mu\left(5\sum\limits_\nu S_\nu S_\nu+7S_m S_m\right), \quad\quad\quad
B_m=6S_m\left(7\sum\limits_\nu S_\nu S_\nu+5S_m S_m\right).
\end{equation}
For simplicity we have here used degenerate light quarks, $u=d=l$. We consider nucleons at zero momentum only and sum over $\mathbf{x}$.

\section{Spectral functions}

In order to determine the nucleon spectral functions, we express the $B$ coefficients in terms of an integral over a kernel and the spectral function. 
For mesons, the spectral relation is given by \cite{Karsch:2003wy,Aarts:2005hg}
\begin{equation}
G(\tau) = \int_{0}^{\infty} \frac{d\omega}{2\pi}\, K(\tau, \omega)\rho(\omega),
\end{equation}
where  the bosonic kernel reads
\begin{equation}
K(\tau, \omega) = \frac{\cosh(\tilde\tau\omega)}{\sinh(\omega/2T)} = [1+n_B(\omega)]e^{-\omega\tau}+n_B(\omega)e^{\omega\tau},
\end{equation}
with $\tilde\tau=\tau-1/2T$ and $n_B(\omega)=1/(e^{\omega/T}-1)$  the Bose distribution.

For baryons the spectral relations are more complicated and read \cite{future} 
\begin{equation}
B_4(\tau)=\int_0^{\infty}\frac{d\omega}{2\pi}\, K_e(\tau,\omega)\rho_4(\omega),
\quad\quad\quad
B_{i,m}(\tau)=\int_0^{\infty}\frac{d\omega}{2\pi}\, K_o(\tau,\omega)\rho_{i,m}(\omega),
\end{equation}
where  $i=1,2,3$. The kernels $K_{e/o}(\tau,\omega)$ are even/odd in $\omega$ respectively and given by
\begin{eqnarray}
&& K_e(\tau,\omega)=\frac{\cosh(\tilde{\tau}\omega)}{\cosh(\omega/2T)}=[1-n_F(\omega)]e^{-\omega\tau}+n_F(\omega)e^{\omega\tau},\\
&& K_o(\tau,\omega)=-\frac{\sinh(\tilde{\tau}\omega)}{\cosh(\omega/2T)}=[1-n_F(\omega)]e^{-\omega\tau}-n_F(\omega)e^{\omega\tau}.
\end{eqnarray}
Here $n_F(\omega)=1/(e^{\omega/T}+1)$ is the Fermi distribution. Note that the normalisation of the kernels is such that $K_{o,e}(\tau,\omega)\to e^{-\omega\tau}$ in the zero-temperature limit.  

Using the parity projector, $\mathrm{P}_{\pm}=\frac{1}{2}(\mathrm{I}_4\pm\gamma_4)$, we construct the positive and negative parity-projected correlation functions,
\begin{equation}
G_{\pm}(\tau)=\int d^3x\; \mathrm{tr}\; \mathrm{P}_{\pm}G(x)=\int d^3x \; \mathrm{tr}\; \langle\mathrm{P}_{\pm} O(x)\bar{O}(0)\rangle=2\left(B_m(\tau)\pm B_4(\tau)\right).
\end{equation}
Hence, $\rho_{\pm}(\omega)=2(\rho_m(\omega)\pm \rho_4(\omega))$. Since $\rho_4(-\omega)=\rho_4(\omega)$ and $\rho_{i,m}(-\omega)=-\rho_{i,m}(\omega)$, we note here that $\rho_{+}(\omega)=-\rho_{-}(-\omega)$.

\section{Contributing processes: Two-loop calculation}

There will be contributions to the spectral function from the decay of a nucleon into 3 quarks with momentum $\mathbf{p}$, $\mathbf{k}$ and $\mathbf{r}$ (where $\mathbf{r}=-\mathbf{p}-\mathbf{k}$), the reverse process and also all possible scattering processes.  At $T=0$ only the first process contributes, while at nonzero temperature all processes are present. The different combinations can be determined by "cutting" the diagram for the nucleon propagator.

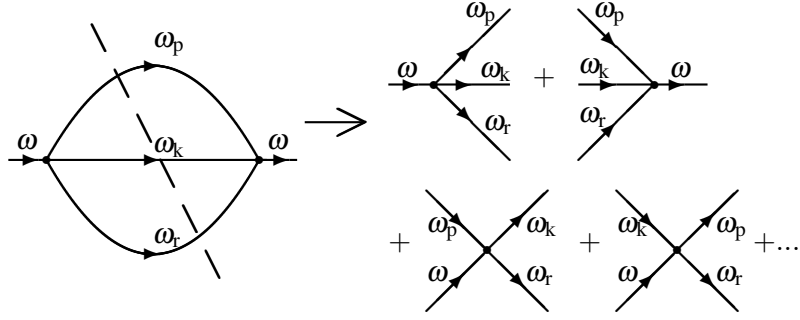
\begin{figure}[h]
\begin{center}

\setlength{\unitlength}{1.cm}
\thicklines

\begin{picture}(10,5.5)
\put(0.5,3){\vector(1,0){1.5}}
\put(0.1,3.15)
{$\mathrm{\omega}$}
\put(3.4,3.15)
{$\mathrm{\omega}$}
\put(1.9,3.1)
{$\mathrm{\omega_k}$}
\put(2.0,3){\line(1,0){1.4}}
\put(0.5,3){\circle*{0.1}}
\put(3.3,3){\circle*{0.1}}
\qbezier(0.5,3)(1.9, 5.5)(3.3, 3)
\qbezier(0.5,3)(1.9, 0.5)(3.3, 3)
\put(1.9,4.25){\vector(1,0){0.1}}
\put(1.9,4.5)
{$\mathrm{\omega_p}$}
\put(1.9,1.75){\vector(1,0){0.1}}
\put(1.9,1.9)
{$\mathrm{\omega_r}$}
\put(0,3){\vector(1,0){0.4}}
\put(0.4,3){\line(1,0){0.1}}
\put(3.3,3){\vector(1,0){0.4}}
\put(3.7,3){\line(1,0){0.1}}
\multiput(1.1,4.8)(0.3,-0.6){6}{\line(1,-2){0.18}}

\put(3.9,3.5){\line(1,0){0.8}}
\put(4.7,3.5){\line(-2,1){0.37}}
\put(4.7,3.5){\line(-2,-1){0.37}}
\put(6.9,4)
{$\mathbf{+}$}

\put(5,4){\vector(1,0){0.4}}
\put(6,4.9)
{$\mathrm{\omega_p}$}
\put(6.25,3.4)
{$\mathrm{\omega_r}$}
\put(6.2,4.1)
{$\mathrm{\omega_k}$}
\put(5.1,4.1)
{$\mathrm{\omega}$}
\put(5.4,4){\line(1,0){0.2}}
\put(5.6,4){\circle*{0.1}}
\put(5.6,4){\vector(1,1){0.5}}
\put(6.1,4.5){\line(1,1){0.5}}
\put(5.6,4){\vector(1,-1){0.5}}
\put(6.1,3.5){\line(1,-1){0.5}}
\put(5.6,4){\vector(1,0){0.5}}
\put(6.1,4){\line(1,0){0.5}}

\put(8.5,4){\vector(1,0){0.4}}
\put(7.7,4.9)
{$\mathrm{\omega_p}$}
\put(7.5,3.5)
{$\mathrm{\omega_r}$}
\put(7.5,4.15)
{$\mathrm{\omega_k}$}
\put(8.7,4.1)
{$\mathrm{\omega}$}
\put(8.9,4){\line(1,0){0.3}}
\put(8.5,4){\circle*{0.1}}
\put(7.5,3){\vector(1,1){0.5}}
\put(8,3.5){\line(1,1){0.5}}
\put(7.5,5){\vector(1,-1){0.5}}
\put(8,4.5){\line(1,-1){0.5}}
\put(7.5,4){\vector(1,0){0.5}}
\put(8,4){\line(1,0){0.5}}

\put(5,1.8)
{$\mathbf{+}$}
\put(5.5,1){\vector(1,1){0.5}}
\put(5.5,1.4)
{$\mathrm{\omega}$}
\put(5.5,2.05)
{$\mathrm{\omega_p}$}
\put(6.8,2.05)
{$\mathrm{\omega_k}$}
\put(6.8,1.4)
{$\mathrm{\omega_r}$}
\put(5.9,1.4){\line(1,1){0.4}}
\put(6.3,1.8){\circle*{0.1}}
\put(6.3,1.8){\vector(1,1){0.5}}
\put(6.7,2.2){\line(1,1){0.4}}
\put(6.3,1.8){\vector(1,-1){0.5}}
\put(6.7,1.4){\line(1,-1){0.4}}
\put(5.5,2.6){\vector(1,-1){0.5}}
\put(5.9,2.2){\line(1,-1){0.4}}

\put(7.5,1.8)
{$\mathbf{+}$}
\put(8,1){\vector(1,1){0.5}}
\put(8,1.4)
{$\mathrm{\omega}$}
\put(8,2.05)
{$\mathrm{\omega_k}$}
\put(9.3,2.05)
{$\mathrm{\omega_p}$}
\put(9.3,1.4)
{$\mathrm{\omega_r}$}
\put(8.4,1.4){\line(1,1){0.4}}
\put(8.8,1.8){\circle*{0.1}}
\put(8.8,1.8){\vector(1,1){0.5}}
\put(9.2,2.2){\line(1,1){0.4}}
\put(8.8,1.8){\vector(1,-1){0.5}}
\put(9.2,1.4){\line(1,-1){0.4}}
\put(8,2.6){\vector(1,-1){0.5}}
\put(8.4,2.2){\line(1,-1){0.4}}

\put(9.8,1.8)
{$\mathbf{+ . . . }$}
\end{picture}
\vspace{-40pt}
\end{center}
\caption{The processes which contribute to the spectral functions are given by $\omega=\pm \omega_\pv\pm \omega_\kv\pm \omega_\rv$.}
\end{figure}

The spectral densities may be written as (with $c=4,i,m$)
\begin{equation}
\rho_c(\omega)=3 \int\limits_{\mathbf{p},\mathbf{k},\mathbf{r}}d\Phi_{\mathbf{p},\mathbf{k},\mathbf{r}}\,\sum\limits_{s_{p,k,r}=\pm1} \mathrm{[stat.]}\,2\pi\delta(\omega+ s_p\omega_\pv +s_k \omega_\kv +s_r \omega_\rv)f_c(\omega, s_p, s_k, s_r, \mathbf{p}, \mathbf{k},\mathbf{r}),
\end{equation}
where
\begin{eqnarray}
&& d\Phi_{\mathbf{p},\mathbf{k},\mathbf{r}}=\frac{d^3p}{(2\pi)^32\omega_\pv}\frac{d^3k}{(2\pi)^32\omega_\kv}\frac{d^3r}{(2\pi)^32\omega_\rv}(2\pi)^3\delta(\mathbf{p}+\mathbf{k}+\mathbf{r}),\qquad\qquad\\
&& \mathrm{[stat.]}=n_F(s_p\omega_\pv)n_F(s_k\omega_\kv)n_F(s_r\omega_\rv)+n_F(-s_p\omega_\pv)n_F(-s_k\omega_\kv)n_F(-s_r\omega_\rv),
\end{eqnarray}
and $f_c$ is dependent on the component $c=4,i,m$.

The integrals can be performed numerically \cite{future}. In the large $\omega$ limit ($\omega\gg T\gg m$) we find
\begin{equation}
\rho_4(\omega)=\frac{5\omega ^5}{2048\pi ^3}\left(1+\frac{112\pi^4 T^4}{3\omega^4}+\dots\right) \label{largelimit1},
\quad\quad 
\rho_m(\omega)=\frac{7m\omega^4}{512\pi^3}\left(1-\frac{4\pi^2T^2}{\omega^2}+\ldots\right),
\end{equation}
where the dots indicate terms that are exponentially suppressed. At zero momentum $\rho_i(\omega)=0$.

\section{On the lattice}

In order to compute the spectral densities on the lattice we perform a two-loop summation over the first Brillouin zone. We consider Wilson fermions, 
see Ref.\ \cite{Aarts:2005hg,future} for more details.
The maximum momenta (and hence energies) are determined by the edges of the Brillouin zones. There are therefore lattice artefacts at large $\omega$.
We also observe lattice artefacts at small $\omega$, which will be discussed further in Ref.\ \cite{future}.
The structure of the spectral densities at large $\omega$ is familiar from similar studies of lattice meson spectral functions
 \cite{Karsch:2003wy,Aarts:2005hg}, although cusps are absent here.

\begin{figure}[h!]
\centering
\includegraphics[width=0.49\textwidth]{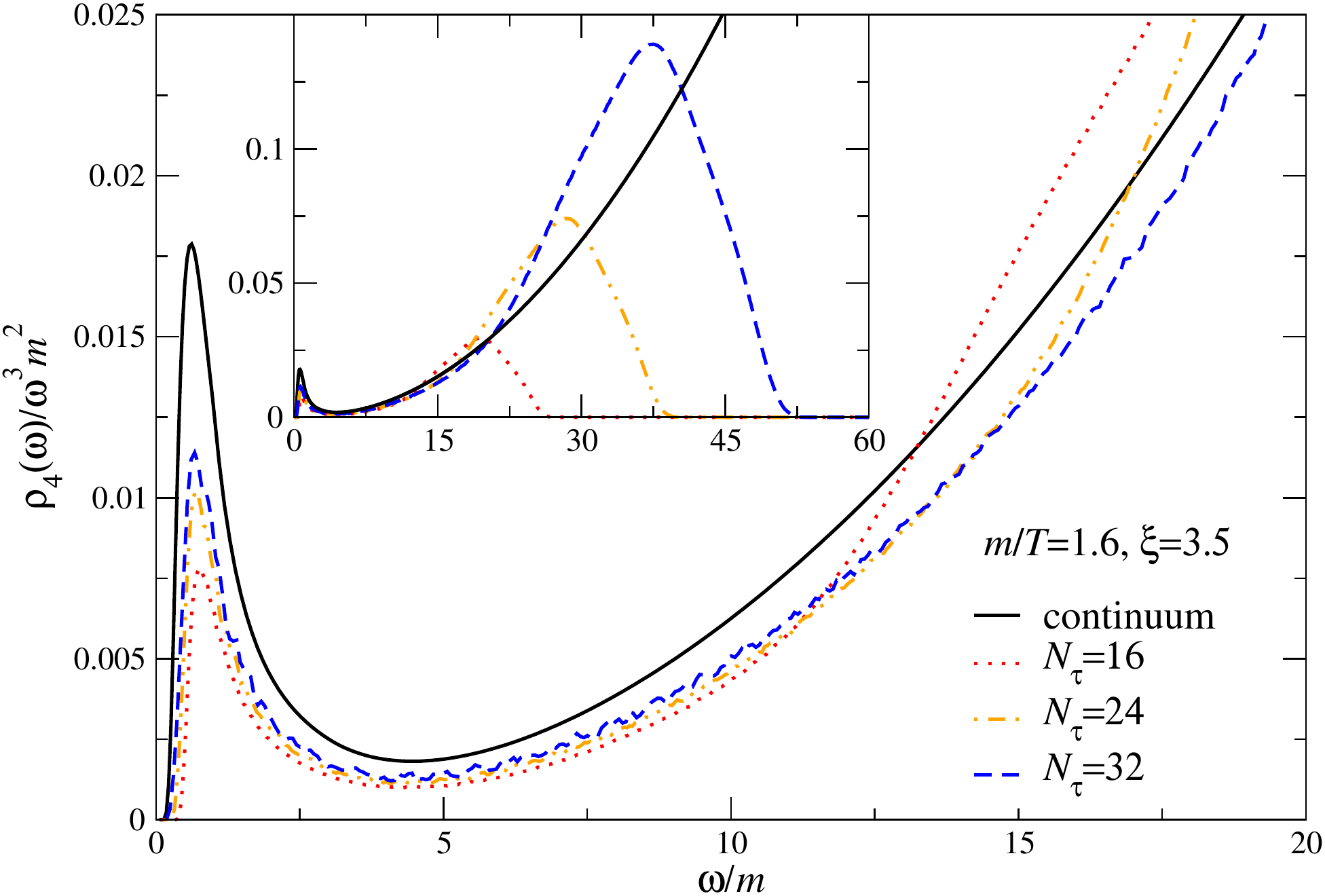} 
\includegraphics[width=0.49\textwidth]{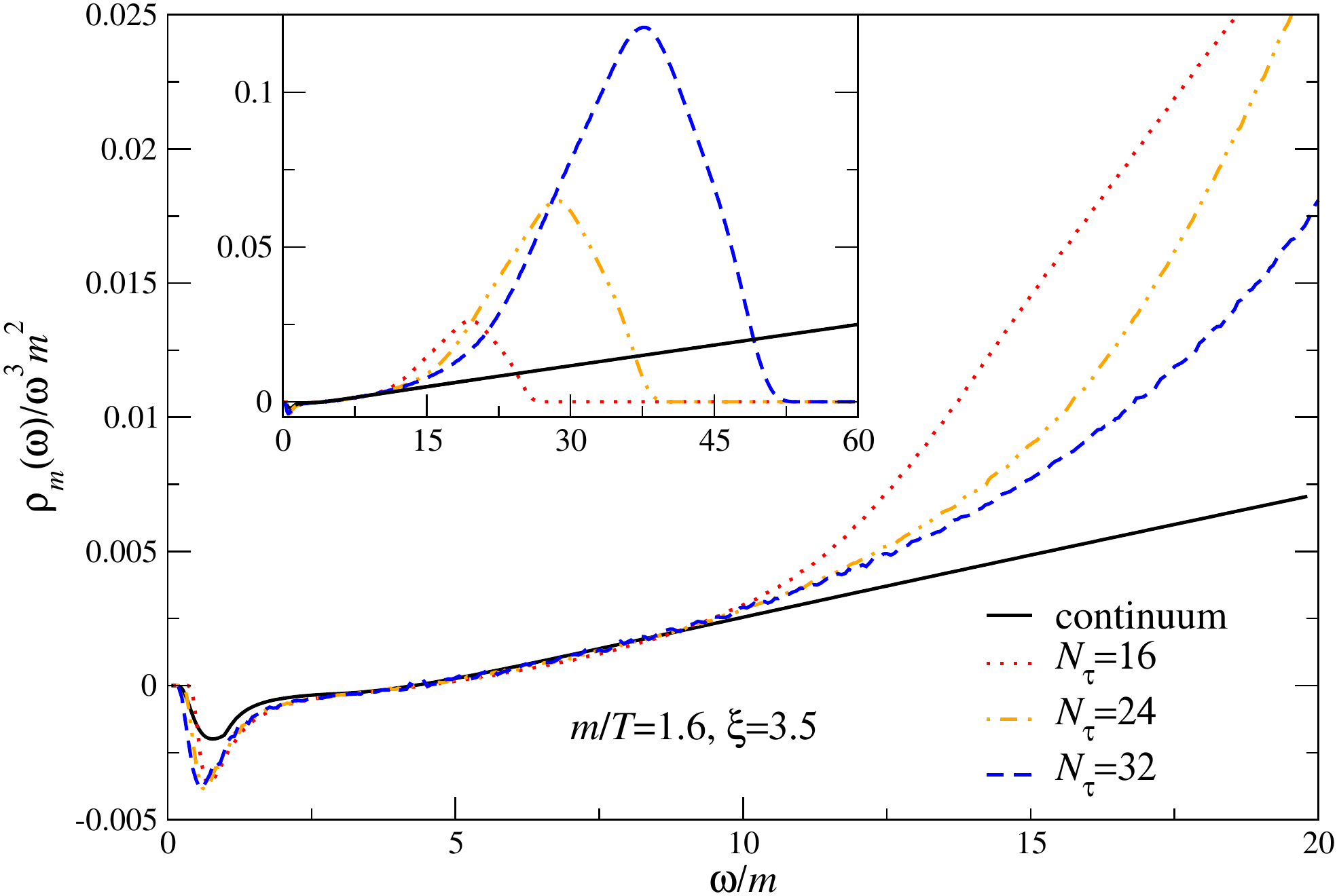}
\caption{Spectral functions $\rho_4(\omega)$ (left) and $\rho_m(\omega)$ (right) scaled with $\omega^3m^2$ in the continuum and for three values of 
$N_\tau$, at fixed anisotropy $\xi=a_s/a_\tau=3.5$ and quark mass $m/T=1.6$. The inset shows the entire domain where the lattice spectral functions are nonzero.
}
\label{w3m2}
 \end{figure}

In the following we present some results, comparing lattice and continuum spectral functions. Following Ref.\ \cite{Aarts:2015mma} we use an anisotropic lattice, with $\xi\equiv a_s/a_\tau\geq 1$. 
In Fig.\ \ref{w3m2} we show the spectral functions $\rho_{4,m}(\omega)$, scaled with $\omega^3m^2$. In the inset the rise at large $\omega$ in the continuum can be seen,  $\rho_4(\omega)/\omega^3\sim \omega^2$ and  $\rho_m(\omega)/\omega^3\sim \omega$, see Eq.\ (\ref{largelimit1}). On the lattice the domain is limited due to the finite Brillouin zone. As $N_\tau$ is increased at fixed $m/T$,  the continuum limit is approached and agreement over a larger range is observed. 
The main frames show a thermal enhancement of both $\rho_4(\omega)$ and $\rho_m(\omega)$ at $\omega\sim m\sim T$. Note that the vacuum contribution only contributes when $\omega>3m$.
While $\rho_4(\omega)$ is positive for all values of $\omega>0$, $\rho_m(\omega)$ is not and has a negative dip at $\omega\sim m$.

\begin{figure}[h]
\centering
\includegraphics[width=0.49\textwidth]{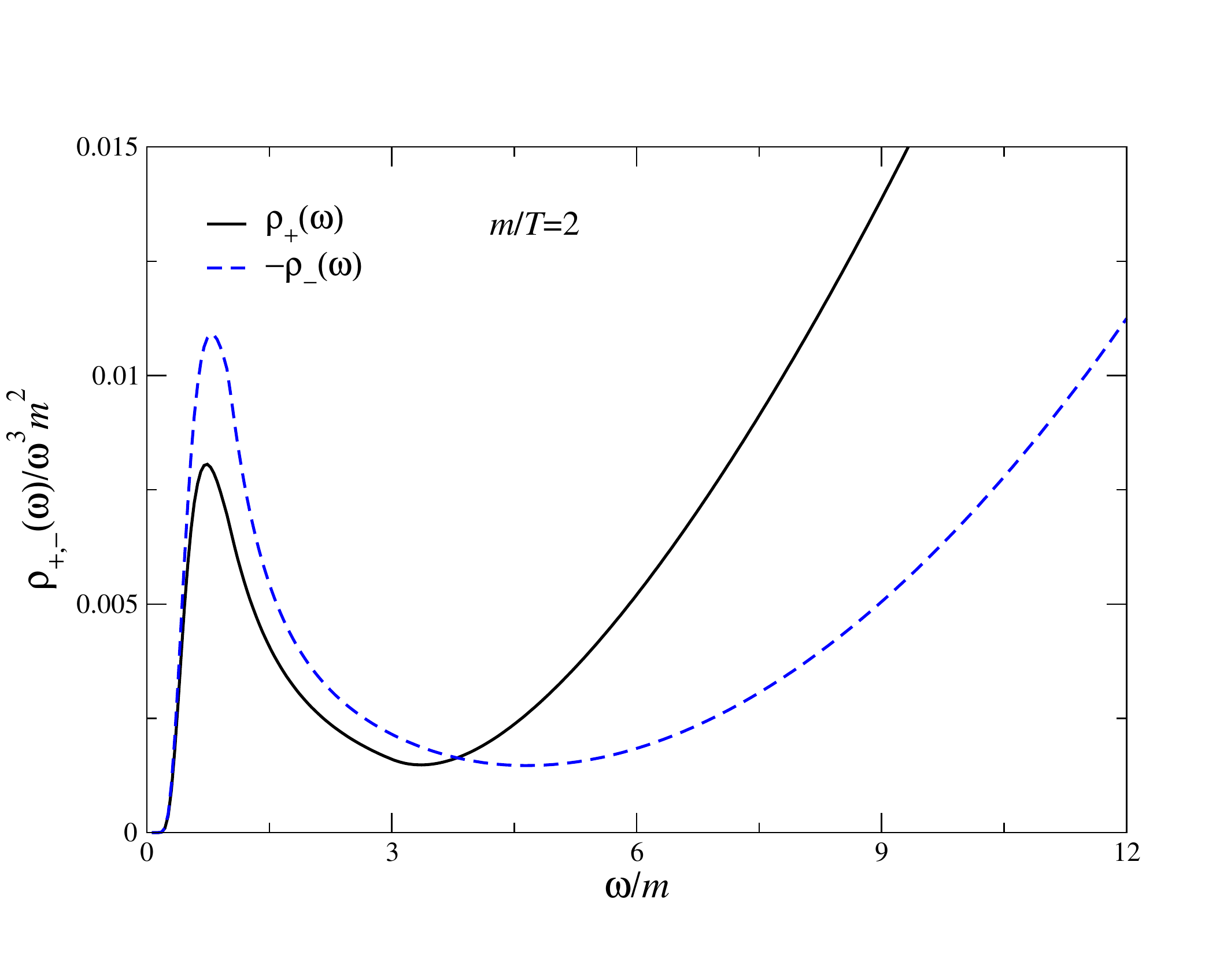}
\includegraphics[width=0.49\textwidth]{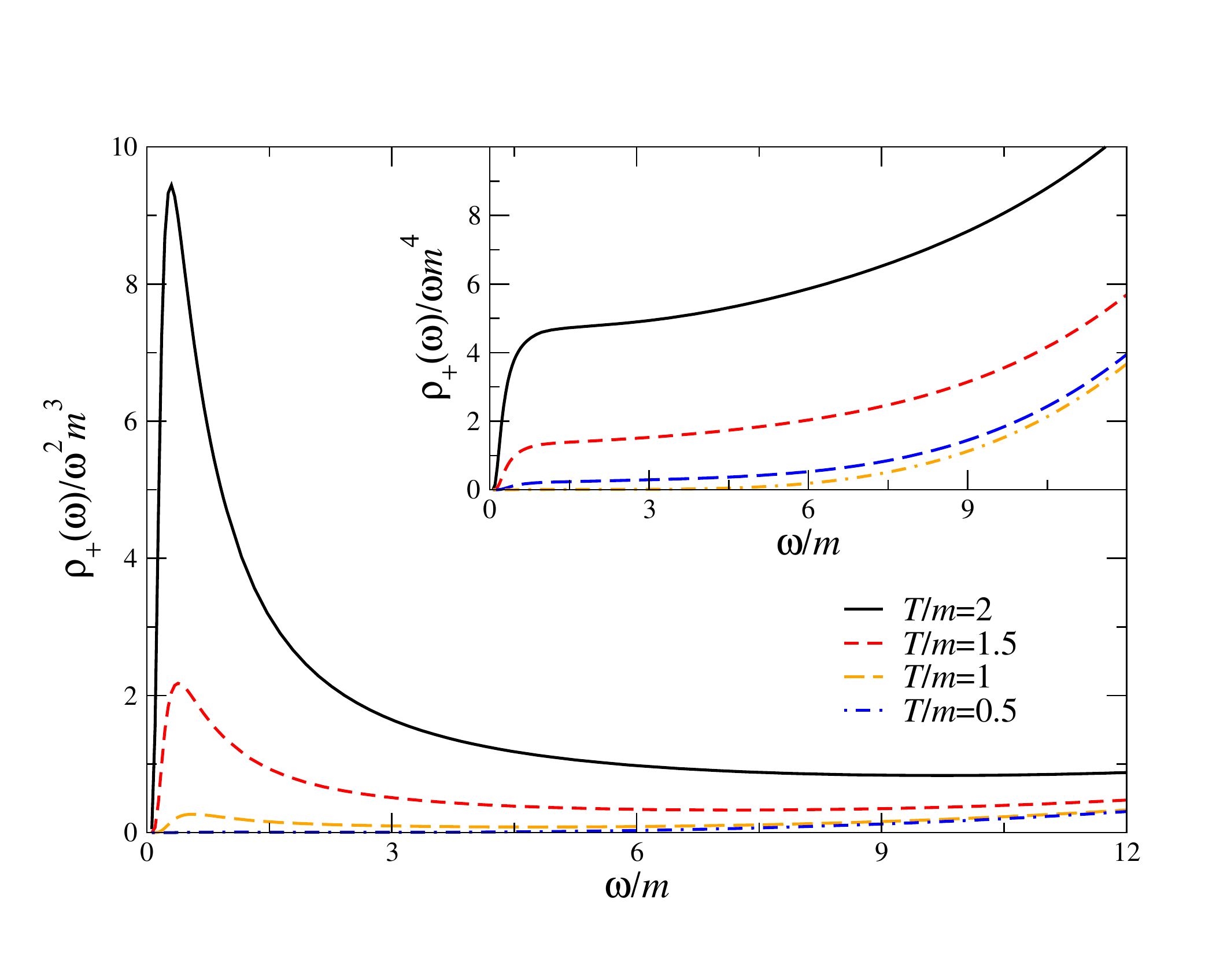}
\caption{Left: Spectral densities $\pm\rho_\pm(\omega)/\omega^3m^2$ with $m/T=2$. Right: $\rho_+(\omega)$ scaled with $\omega^2m^3$  and $\omega m^4$ (inset) for four values of $T/m$.
}
\label{rhopm}
\end{figure}

In Fig.\ \ref{rhopm} (left) $\pm\rho_\pm(\omega)$ are shown. We note that  $\rho_+(\omega)>-\rho_-(\omega)$ for large $\omega$. This  is expected since $\pm\rho_\pm(\omega)=2(\rho_4(\omega)\pm \rho_m(\omega))$. Furthermore we observe that  $\pm\rho_\pm(\omega)>0$ for all $\omega>0$, even though $\rho_m(\omega)$ is not. 
Fig.\ \ref{rhopm} (right) shows $\rho_+(\omega)$ at four temperatures. We see a reduction of the thermal enhancement as the temperature is decreased.
The inset shows $\rho_+(\omega)$ scaled with $\omega$, to demonstrate that the apparent peaks depend on the choice of normalisation and do not correspond to a physical particle (bound state) in the spectrum.

\section{Chiral symmetry, varying the anisotropy}

\begin{figure}[h]
 \centering
\includegraphics[width=0.49\textwidth]{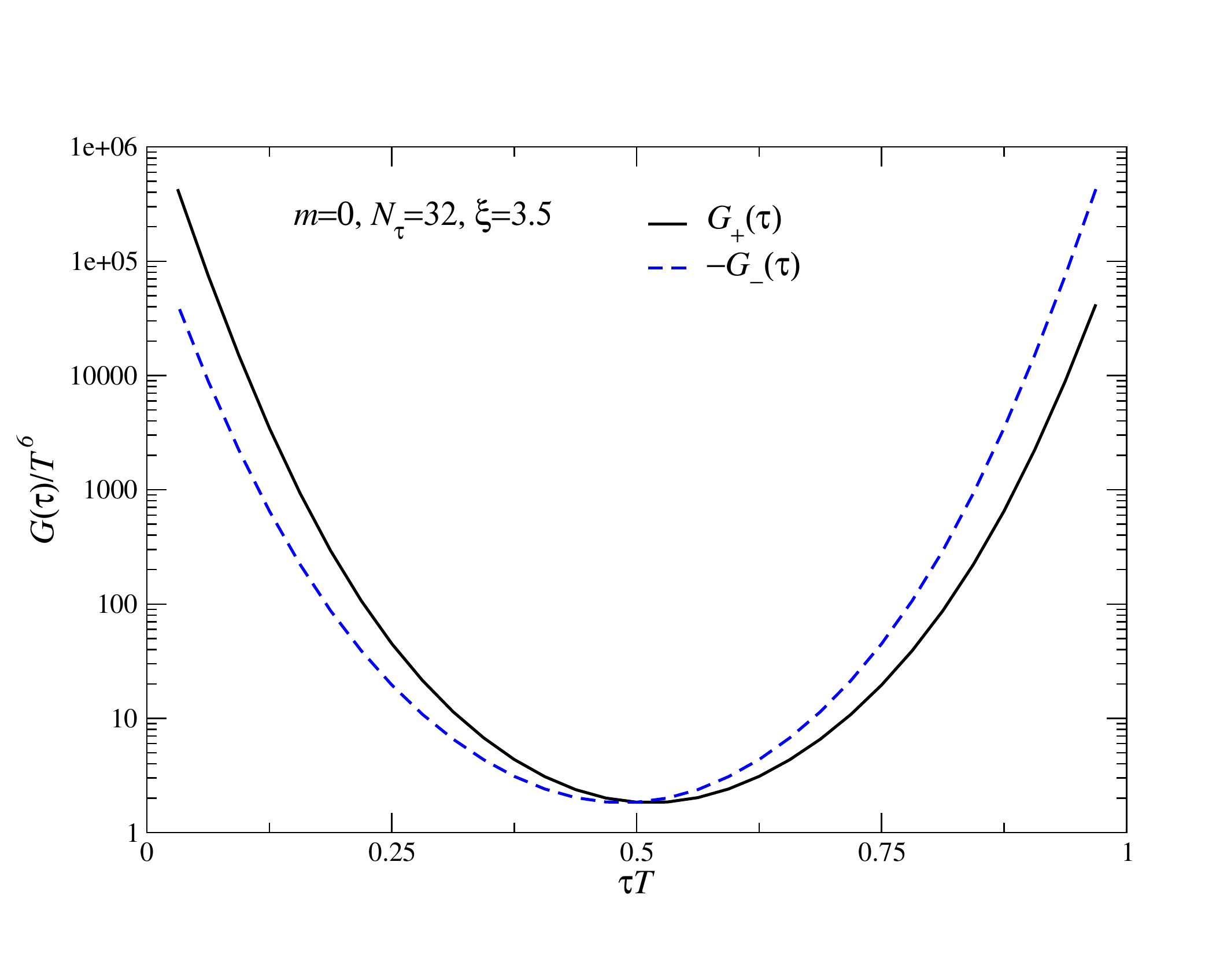}
\includegraphics[width=0.49\textwidth]{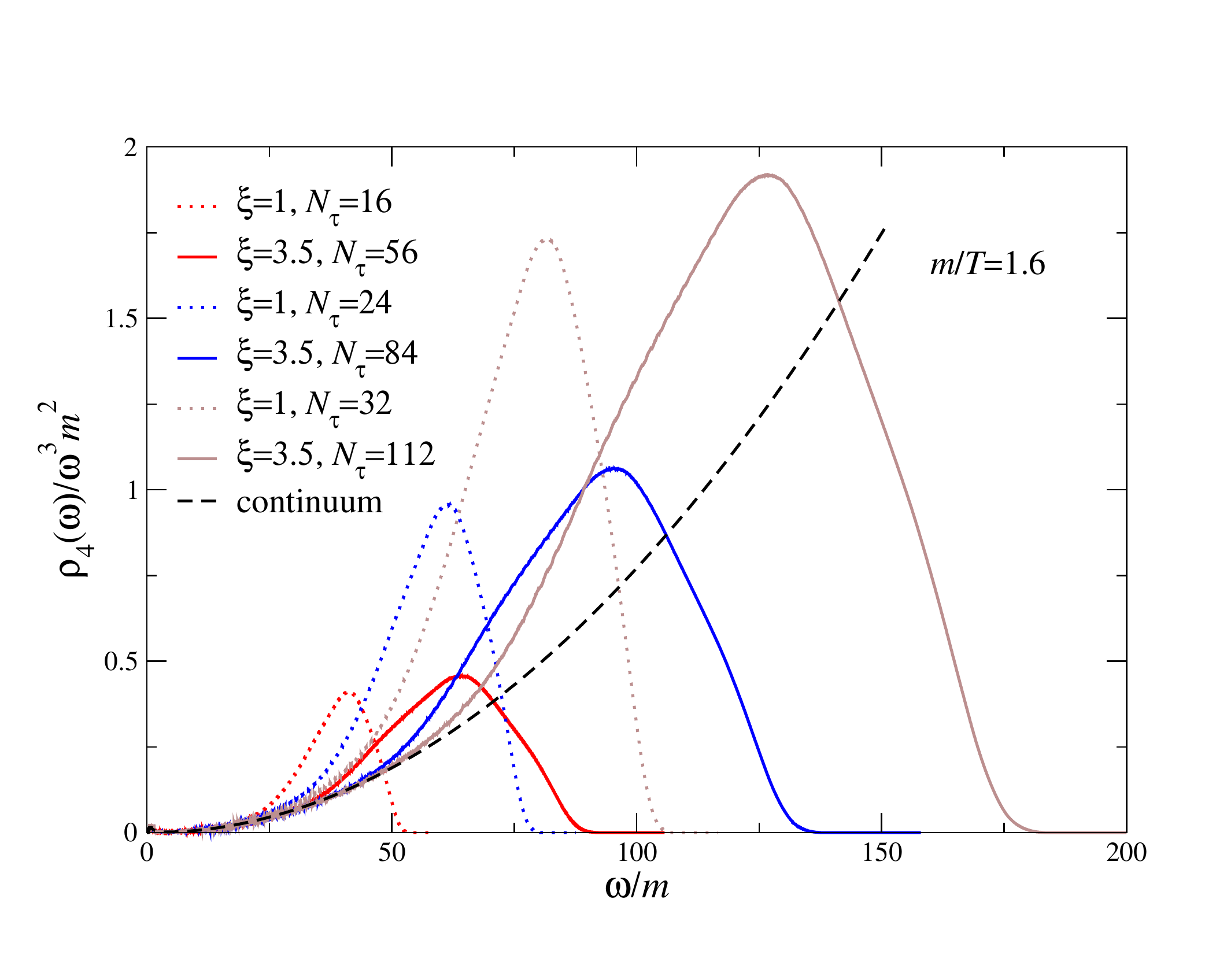}
\caption{Left: Euclidean correlators $\pm G_\pm(\tau)/T^6$ in the massless case for Wilson fermions. 
Right: Spectral densities $\rho_4(\omega)/\omega^3m^2$ at fixed $m/T=1.6$ for different values of the lattice spacing $a_sT=\xi/N_\tau$, each at two values of $\xi$.
 }
\label{chiral}
 \end{figure}

In the massless case we find that in the continuum $\rho_m(\omega)=0$ and hence $G_+(\tau)=-G_-(\tau)=G_+(1/T-\tau)=2B_4(\tau)$, which implies there is parity doubling \cite{Aarts:2015mma}.
However, on the lattice, the Wilson term breaks chiral symmetry at short distance. This means that even for vanishing mass parameter,  $\rho_m(\omega)\ne 0$, and $G_+(\tau)\ne -G_-(\tau)$. This is demonstrated in Fig.\ \ref{chiral} (left), where the nondegeneracy of $G_\pm(\tau)$ (and the resulting absence of the parity doubling) is shown. This plot should be compared with the results in the interacting case \cite{Aarts:2015mma}, where the correlators are considerably more skewed below $T_c$, due to the spontaneous breaking of chiral symmetry at low energies.

The lattice artefacts at large $\omega$ depend on the lattice cutoff. This is demonstrated in Fig.\ \ref{chiral} (right), where $\rho_4(\omega)$ is shown for three different values of the spatial lattice spacing $a_s$. Increasing both $N_\tau$ and $\xi$ shifts the maximal energy to larger values, similar to the mesonic case \cite{Aarts:2005hg}.

\section{Conclusion}

The perturbative calculation of baryon spectral densities involves a two-loop calculation, already at lowest order in the strong-coupling constant.  Hence it is more involved compared to mesonic spectral functions, where the leading behaviour is determined by a one-loop computation.  
We have seen that at nonzero temperature, there is clear thermal enhancement at $\omega \sim T\sim m$, which, depending on the normalisation, may take the form of a peak in the spectral function.

On the lattice there are lattice artefacts arising from the sum over finite Brillouin zones, both at large and small values of $\omega$.
Since we have used Wilson lattice fermions, chiral symmetry is broken at short distances, leading to a nondegeneracy in the two channels related by parity. This effect is, however, much smaller than the effect of spontaneous chiral symmetry breaking at low temperatures in the interacting case.
 The next step will be the calculation of nucleon spectral functions for the interacting case \cite{Allton:lat15}, using the correlators previously analysed in Ref.\ 
 \cite{Aarts:2015mma}.

\acknowledgments

We thank Chris Allton, Simon Hands, Benjamin J\"ager and Jonivar Skullerud for collaboration on Ref.\ \cite{Aarts:2015mma}.
This work is supported by STFC via grant ST/L000369/1, the Royal Society and the Wolfson Foundation.

\end{document}